\documentclass[12pt]{article}

\usepackage{amsfonts}
\usepackage{mathrsfs}
\usepackage{amsmath,amssymb,amsthm}
\usepackage{mathrsfs}
\usepackage{latexsym}
\usepackage{indentfirst}
\usepackage[top=1in, bottom=1in, left=1in, right=1in]{geometry}
\usepackage{color}
\usepackage{footmisc}
\usepackage{endnotes}
\usepackage{float}
\usepackage{graphicx}
\usepackage{lscape}

\usepackage[english]{babel}
\usepackage[utf8]{inputenc}
\usepackage{algorithm}
\usepackage[noend]{algpseudocode}

\usepackage{appendix}

\begin{document}

{\Large \bf Combining Random Walks and Nonparametric Bayesian Topic Model for Community Detection}

Ruimin Zhu\footnote{Department of Statistics, Northwestern University, ruiminzhu2014@u.northwestern.edu} and Wenxin Jiang\footnote{Department of Statistics, Northwestern University, wjiang@northwestern.edu} \\

\today\\

\textbf{Abstract} Community detection has been an active research area for decades. Among all probabilistic models, Stochastic Block Model has been the most popular one. This paper introduces a novel probabilistic model: RW-HDP, based on random walks and Hierarchical Dirichlet Process, for community extraction. In RW-HDP, random walks conducted in a social network are treated as documents; nodes are treated as words. By using Hierarchical Dirichlet Process, a nonparametric Bayesian model, we are not only able to cluster nodes into different communities, but also determine the number of communities automatically. We use Stochastic Variational Inference for our model inference, which makes our method time efficient and can be easily extended to an online learning algorithm. \\


\section{Introduction}
Social networks have been studied intensively for decades. Community detection is one of the most interesting problems in this area. Social networks usually possess community structure: nodes in the network can be grouped such that nodes in the same group are densely connected while nodes in different groups are sparsely connected. We call such groups communities. \\

Over the years, many novel methods have been proposed for community structure extraction. Among them, there is a category of methods based on random walks on networks \cite{deepwalk}, \cite{walktrap}, \cite{maps}. In these methods, random walks on the network serve as input data for future network property mining. A simple illustration explains why random walks give us insights of the network's properties, such as community structure. Because nodes in the same community tend to be densely connected, unsurprisingly, the path of a random walker tends to fall in only a small number of communities. Combining many random walks together can be a very good way of collecting information of the network, thus reveal the network's properties. What's more, a random walker chooses its next move based solely on the information of the neighborhood of its current position, which makes random walk easy to implement. \\

Traditional community detection methods need to specify the number of communities manually. Moreover, this same problem is quite common in statistical and machine learning. How many classes should I use in my mixture model? How many factors should I use in my factor analysis \cite{bnp-tutorial}? The classical way to handle this problem is by model selection, where people first fit several models, then conduct model selection to choose the model that fits the data well and also has a relatively small model complexity. Nonparametric Bayesian models (BNP) address this problem in a very different way: it can detect such hyper-parameters automatically and can increase the model complexity when new data are observed. \\

Our model RW-HDP is based on random walks and nonparametric Bayesian modeling mentioned above. We first conduct random walks on the network. Each node is treated as a word, and each random walk is treated as a document. Using random walks as the input data, we fit the HDP topic model to reveal community structure. For fast model inference (posterior approximation), we use Stochastic Variational Inference \cite{svi}. \\

Our model has many merits:
\begin{itemize}
	\item By equipping a probabilistic model to the random walks, we obtain a better interpretation than other methods.
	\item HDP, a nonparametric Bayesian model, can automatically determine the number of communities and, if necessary, automatically increase the number of communities when new data are observed.
	\item The random walk part can be parallelized which ensures that our method can be time efficient.
	\item The posterior approximation part can easily be extended to an online learning scenario where random walks are not generated at once but keep streaming.
\end{itemize}

The rest of the paper is organized as follows. In section 2 we introduce the related works. In section 3, we describe our method in details. In section 4, we outline the experimental studies and compare our method to others. In section 5, we close with a conclusion and a discussion of possible future works.

\section{Background and Related Works}
Community detection in complex networks has been receiving a lot of attentions during the past decades. A large number of methods have been proposed for this problem. For example, centrality or betweenness based approaches and minimum cut methods are two of the most popular and rich tactics \cite{freeman}, \cite{girvan}, \cite{wilkinson}, \cite{garey}, \cite{flake}, \cite{tarjan}. \\

While being less popular, other works such as random walk based methods also enrich the research in this field. Pons and Latapy \cite{walktrap} introduced a measure of similarity between nodes based on random walks. Rosvall and Bergstrom \cite{maps} utilized the probability flow of random walks on a network as a proxy for information flows to reveal community structure. Perozzi, AI-Rfou, and Skiena \cite{deepwalk} conducted random walks on a network, and treated them as sentences, later they used Natural Language Processing techniques to map the nodes into a Euclidean space for various tasks including clustering analysis. \\

LDA topic model proposed by Blei, et al \cite{lda} has been applied in many domains such as document modeling \cite{lda}, image processing \cite{sudderth}, information retrieval \cite{wei} and the list goes on. Zhang, et al \cite{sip-lda} developed a method for network community detection based on LDA. In their model, a social interaction profile was treated as a document and LDA was smoothly fitted into the community detection task, where communities were treated as topics. \\

Nonparametric Bayesian models are becoming popular these days due to their flexibility \cite{bnp-tutorial}, \cite{hjort} and new posterior inference methods such as Variational Inference \cite{vi-graph}, \cite{wainwright}. Teh, Jordan, Beal, and Blei \cite{hdp} extended the Dirichlet Process to the Hierarchical Dirichlet Process which can also be viewed as an extension of LDA model. Hoffman, Blei, Wang, and Paisley \cite{svi} developed the Stochastic Variational Inference for several non-parametric Bayesian models including HDP. In social network research, Morup and Schmidt \cite{bcd}, \cite{mikkel} formulated a non-parametric Bayesian community generative model for social network analysis. Kim, Gopalan, Blei, and Sudderth \cite{sudderth} proposed the hierarchical Dirichlet process relational model which allowed nodes to have mixed membership in an unbounded set of communities. Guo and Nordman \cite{guo} introduced a series of Bayesian nonparametric statistical models for community detection. Blundell and Teh \cite{blundell} proposed an efficient Bayesian nonparametric model for discovering hierarchical community structure in social networks. \\

In relation to these related works, our work essentially combines the data generation component of the random walk (RW) approach and the inferential component of the HDP topic model.  This combination allows us to achieve more than what the previous methods can do, e.g., automatic detection of number of communities, and the use of more informative RW ``documents'' for the topic model to detect the network communities. \\

Later we will show that the proposed method indeed leads to obvious improvement in several performance measures in the real applications, when compared to several existing methods; but first we will describe the details of the proposed method in the next section.
\section{Random Walk Hierarchical Dirichlet Process Topic Model}
In this section, we describe the RW-HDP model for network community detection in detail. We first introduce related terminologies and notation. Section 3.1 describes the corpus generation using random walks. Section 3.2 introduces the Hierarchical Dirichlet Process topic model. \\

Formally, let $\mathscr{G=(V, E)}$ be an undirected network, where $\mathscr{V}$ is the set of nodes; $\mathscr{E}=\{e_{ij}\}$ is the set of edges and $e_{ij}$ is the weight between node $i$ and $j$. $V=|\mathscr{V}|$ is the number of nodes. As we treat each node as a word, $V$ is also the vocabulary size. We assume networks are undirected but our model can easily be extended to directed ones. \\

Our notation is partly summarized in Table 1. 
\begin{table}[ht]
\centering
\begin{tabular}{l|l|l}
	\hline
	$V$ & number of nodes & size of the vocabulary \\
	$D$ & number of random walks & number of documents \\
	$L$ & expected random walk length & expected document length \\
	$N_d$ & length of the $d^{th}$ random walk & length of the $d^{th}$ document \\
	$w_d$ & the $d^{th}$ random walk & the $d^{th}$ document \\
	$w_{dn}$ & $n^{th}$ node in the $d^{th}$ random walk & $n^{th}$ word in the $d^{th}$ document \\
	$z_{dn}$ & community assignment of $w_{dn}$ & topic assignment of $w_{dn}$ \\
	\hline
\end{tabular}
\caption{Notation}
\label{table:notation}
\end{table}

\subsection{Random Walk for Corpus Generation}
Each random walk is treated as a document and the collection of random walks is treated as the corpus. To carry out a random walk, we need to specify three elements: the starting point, the length, and the transition probability matrix. \\

The starting points are sampled independently uniformly across the nodes set $\mathscr{V}$. The lengths of random walks are sampled independently from Poisson distribution parametrized by $L$, the expected length of a random walk. As for the transition matrix $P = (p_{ij})_{V\times V}$, we take the edge weights into consideration. Let $p_{ij}=\frac{e_{ij}}{\Sigma_j e_{ij}}$. That being said, a random walker at node $i$, would randomly choose one of node $i$'s neighbors to visit in the next step with probability proportional to the edge weight between node $i$ and its destination. \\

The success of the proposed method is caused by using RW to generate informative ``documents'' for topic discovery. The same word, when embedded in good documents, can be easily assigned its correct topic; on the other hand, when it is embedded in bad documents, it cannot be easily assigned any topic with clarity, even when the same topic model, e.g., the HDP topic model, is used. As a simple example: when the network has two communities of integer-indexed nodes, the community of odd indexes and the community of even indexes, with many more within-community connections than between-community connections. The completely random (CR) documents would have documents all looking like $\{1,4,3,2,...\}$, which is hard to be assigned any topic clearly. On the other hand, RW will generate documents either like $\{2,8,4,10,...\}$ or $\{3,11,9,5,...\}$,which have clearly distinctive styles of word distributions. The same word, say, 3, cannot be easily assigned a topic in CR documents, but can be easily assigned a topic in the RW documents. \\

The RW documents also have a natural interpretation. An RW document can be regarded as recording the experience of a person in making a sequence of friends in an RW fashion in the network, each time with his newest friend introducing a ``neighbor'' to him as a newer friend. These documents of names of friends, when accumulated over many such experiences as a corpus will, therefore, be very informative in inferring the community structure, unlike CR documents consisting of only random names. \\

To speed up the corpus generation process, random walks can be carried out simultaneously using parallel computing. Also, if we keep carrying out random walks and feeding newly observed random walks to the HDP topic model, we easily extend our model to an online learning model. Although we set the length of a random walk be a Poisson random variable, it is not critical to our method.

\subsection{Hierarchical Dirichlet Process Topic Model}
In this section, we introduce the Hierarchical Dirichlet Process topic model \cite{hdp} in detail and show how to apply it for network community structure extraction. \\

HDP topic model is a mixture model with unbounded number of mixtures (topics). In mixture models, each observation is assumed to belong to a cluster or group. In our case, each node $w_{dn}$ belongs to a topic $z_{dn}$. Note that in topic models, a word may be assigned to different topics in different documents. For the network community detection problem, we go a step further: we use Bayesian rule to calculate the overall probability of a node belonging to each community and assign the node to the community with the largest posterior probability. If we can get the conditional probability $p(w|z)$ and the topic probability $p(z)$, then by Bayesian rule, the conditional probability $p(z|w)$ is given by
\begin{equation}
p(z|w) = \frac{p(w|z)p(z)}{p(w)} \propto p(w|z)p(z).
\label{eq: inference}
\end{equation}
The community assignment of node $w$ is $\underset{k \in \{1,2,\dots\}}{\operatorname{argmax}}p(z=k|w)$. 

For programming implementation simplicity, we will derive the HDP topic model using stick-breaking construction. The HDP topic model couples a set of document-level Dirichlet Process via a single top-level Dirichlet Process. In the top-level, the base measure $H$ is a symmetric Dirichlet over the vocabulary simplex. The stick-breaking construction for HDP topic model is given by \cite{svi}:
\begin{enumerate}
	\item Draw an infinite number of topics, $\phi_k\sim Dirichlet(\eta), k = 1, 2, \dots$
	\item Draw corpus breaking proportions, $v_k\sim Beta(1, \gamma), k=1, 2, \dots$
	\item For each document:
	\begin{enumerate}
		\item Draw document-level topic indexes, $c_{di}\sim Multinomial(\sigma(v)), i = 1, 2, \dots$
		\item Draw document breaking proportions, $\pi_{di}\sim Beta(1, \alpha), i = 1, 2, \dots$
		\item For each word:
		\begin{enumerate}
			\item Draw topic assignment $z_{dn}\sim Multinomial(\sigma(\pi_d))$.
			\item Draw word $w_{dn}\sim Multinomial(\phi_{c_{d, z_{dn}}})$.
		\end{enumerate}
	\end{enumerate}
\end{enumerate}

Note that $\phi_k = p(w|z=k) = [p(w_1|z=k), \dots, p(w_V|z=k)]$ specifies the word distribution in the $k^{th}$ topic. By stick-construction, the corpus level distribution of topics is given by:
\begin{align*}
& p(z=1) = \sigma_1(v) = v_1, \\
& p(z=k) = \sigma_k(v) = v_k(1-\sum_{j=1}^{k-1}\sigma_j(v)), k = 2, 3, \dots
\end{align*}
thus from (1) the overall probability of node $i$ belonging to topic $k$ is given by
\begin{equation}
p(z=k|w_i) \propto p(w_i|k)p(z=k) = \phi_{ki}v_k(1-\sum_{j=1}^{k-1}\sigma_j(v)).
\end{equation}

The HDP topic model is a hierarchical model which means that there are different levels in the governing ranges of variables. Variables such as $\phi_k$ and $v_k$ are corpus or global level variables. They govern the distributions of the observations (words) across all documents. While, other variables such as $z_{dn}$ and $\pi_{di}$ are document or local level variables because they only govern the distributions of the observations in a particular document. The scopes of variables are clearly shown in Figure 1.
\begin{figure}[H]
	\centering
	\includegraphics[scale=0.6]{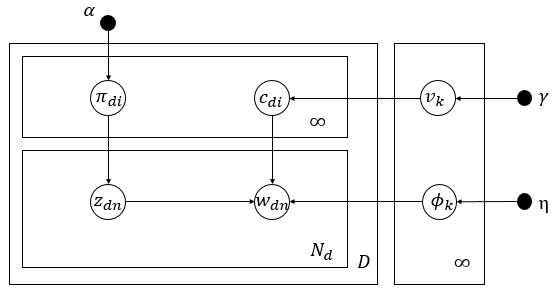}
	\caption{HDP Topic Model}
\end{figure}

We choose Stochastic Variational Inference \cite{svi}, which is a combination of stochastic method and Variational Inference, for posterior approximation. At each iteration, we only sample a mini batch of documents, update their corresponding local variables, and then treat them as if they appear $D$ times to update the global variables\footnote{Other constructions of the HDP topic model and the details of the Stochastic Variational Inference for the model can be found in Hoffman et al. (2013).}.

\section{Experiments}
In this section, we evaluate and compare the performances of our model with three previously proposed methods. The comparison has been done in five different types of networks. 
 
\subsection{Models}
We deliberately choose three models to compare with ours. Each of them shares one characteristic with our model. Since RW-HDP consists of three major characteristics: random walk, nonparametric Bayesian, and topic model. We decide to choose the following three models to compare with:
\begin{enumerate}
	\item SIP2-LDA \cite{sip-lda}, a topic based community detection model. In this model, each node is associated with a social interaction profile (SIP2), which only takes a node's immediate and secondary neighbors into consideration. Those social interaction profiles are treated as documents for community detection using Latent Dirichlet Allocation.	
	\item Walktrap \cite{walktrap}, a random walk based community detection model. This method does not actually implement random walks on the network, but it defines node-to-node distance and community-to-community distance based on properties of random walks, such as the transition probability between any pair of nodes within $t$ steps. Later, it merges communities iteratively to get a hierarchical tree of partition. Finally, it cuts the tree to get the best partition.
	\item BCD \cite{bcd}, a nonparametric Bayesian network generative model. The generative process is: first, a cluster assignment is generated using Chinese Restaurant Process (a commonly used metaphor for Dirichlet Process); then, within-cluster and between-cluster link probabilities are generated; finally, links between nodes are generated according to the within- and between-cluster link probabilities.
\end{enumerate} 

\subsection{Data}
We conduct experiments on several different types of networks:
\begin{enumerate}
	\item yeast: a yeast protein complex interaction network \cite{yeast}.
	\item GSE: a breast cancer gene co-expression network \cite{gse1}, \cite{gse2}.
	\item ca-GrQc: Arxiv General Relativity and Quantum Cosmology collaboration network. If an author $i$ co-authored a paper with author $j$, the graph contains an undirected edge between $i$ and $j$ \cite{grqc}.
	\item ca-CondMat: Arxiv Condense Matter Physics collaboration network \cite{grqc}.
	\item US powergrid: the high-voltage power grid in the Western States of the United States of America. The nodes are transformers, substations, and generators, and the ties are high-voltage transmission lines \cite{powergrid}.
\end{enumerate}

The statistics of networks are summarized in Table 2.
\begin{table}[ht]
	\centering
	\begin{tabular}{l|c|c|c|c|c}
		\hline
		statistics & yeast & GSE & ca-GrQc & ca-CondMat & US powergrid \\
		type & biology & biology & co-authorship & co-authorship & engineer \\
		nodes & 1540 & 9112 & 5242 & 16264 & 4941 \\
		edges & 8703 & 244928 & 14478 & 47594 & 6594 \\
		\hline
	\end{tabular}
	\caption{Network Statistics}
	\label{table:networks}
\end{table}

\subsection{Evaluation Metrics}
To compare our method with others, we use community scoring functions. A good community tends to be densely connected internally and sparsely connected with other parts of the network. The community scoring functions quantify this intuition in different aspects \cite{metrics}. \\

Given a set of nodes $S$(a community of the network $\mathscr{G=(V, E)}$), let $(S, S(\mathscr{E}))$ be the subgraph induced by $S$, where $S(\mathscr{E}) = \{(u, v)\in \mathscr{E}:u\in S, v\in S\}$. Let $n$ denote the number of nodes in the original network. Let $n_S$ denote the number of nodes in set $S$, and $m_S$ the number of edges in set $S(\mathscr{E})$. Let $c_S$ denote the number of edges with one end in $S$, and the other outside of $S$: $c_S = |\{(u, v):u\in S, v\notin S\}|$. We use the following four community scoring functions:
\begin{enumerate}
	\item Internal density: $D=\frac{2m_S}{n_S(n_S-1)}$. This metric sores the community structure based on its internal connectivity. A larger internal density usually means a better community structure \cite{density}.
	\item Cut Ratio: $CR=\frac{c_S}{n_S(n-n_S)}$, which quantifies the community structure based on its external connectivity. A smaller cut ratio usually means a better community structure \cite{cut}.
	\item Conductance: $C=\frac{c_S}{2m_S+c_S}$, which measures the fraction of edge that points outside the cluster. It combines both internal and external connectivity to give a score. A smaller conductance usually means a better community structure \cite{conductance}.
	\item Modularity: $Q=\sum_{i=1}^{m}(e_{ii}-a_i^2)$, where $m$ is the number of communities, $e_{ij}$ the fraction of edges with one end in community $i$ and the other in community $j$, $a_i =\sum_je_{ij}$. This index falls in $[-0.5, 1)$. A larger modularity means a better community structure \cite{modularity}. 
\end{enumerate}

For the two topic based models: SIP2-LDA and RW-HDP, we also compare their perplexity scores \cite{lda} on the testing corpus. Perplexity is defined as:
$$
perplexity(D_{test}) = exp^{-\frac{\sum_{d=1}^{M}logp(w_d)}{\sum_{d=1}^{M}N_d}},
$$
where $D_{test}$ is the testing corpus, which is either random walks in RW-HDP model, or randomly sampled social profiles in SIP-LDA model. A smaller perplexity score corresponds to a better topic model.

\subsection{Choice of Hyperparameters}
There are three types of hyperparameters that are worthy of study. They are corpus, HDP topic model, and Stochastic Variational Inference hyperparameters. Here, we briefly explain the principles and intuitions of choosing proper hyperparameters. \\

The corpus hyperparameters are average random walk length $L$ and number of random walks $D$. If $L$ is too large, a document will contain words from many topics, which by intuition is not a well-written document and thus will make it hard for the topic model to detect any meaningful topic assignment. $L$ cannot be too small either. An extreme case is $L=1$, where it is impossible to capture the dependencies between nodes and thus will make community detection by topic models in vain. In our experiments, we set $L$ to be around 100. The same principle applies to corpus size $D$. We set $D$ to be approximately five times of the number of nodes. \\

The HDP topic model has an infinite number of topics which makes it hard for programming implementation and variational inference. Instead, we do truncations both at the corpus level and the document level. At the corpus level, we fit $K$ breaking points as the topic choices. At the document level, we fit $T$ topic pointers and let the topic assignment variable take on one of $T$ values. $T$ can be much smaller than $K$ as there might be hundreds of topics in the corpus but for a single document, only a small number of topics will be exhibited. To see this is still an infinite model, note that by setting the truncations high enough, the variational posterior will not necessarily use all topics \cite{svi}. \\

The last type of hyperparameters is those in Stochastic Variational Inference, such as batch size, the number of epochs, etc. The choice of batch size is a tradeoff between speed and noise. We set the batch size to be a small number such as $2$ since noise is tolerable in our community detection task. We run $3$ epochs on the corpus.

\subsection{Results and Comparison}
\begin{table}[H]
	\centering
	\begin{tabular}{l|c|c|c|c|c}
		\hline
		model & yeast & GSE & ca-GrQc & ca-CondMat & US powergrid \\
		\hline
		RW-HDP & 0.7605 & 0.5967 & 0.7848 & 0.7588 & 0.9087 \\
		SIP2-LDA & 0.6995 & 0.5881 & 0.7479 & 0.6615 & 0.7775 \\
		Walktrap & 0.6968 & 0.6014 & 0.7430 & 0.7238 & 0.8953 \\
		BCD & 0.6452 & 0.2017 & 0.5378 & 0.5041 & 0.4802 \\
		\hline
	\end{tabular}
	\caption{modularity}
	\label{table:modularity}
\end{table}

\begin{table}[H]
	\centering
	\begin{tabular}{l|c|c|c|c|c}
		\hline
		model & yeast & GSE & ca-GrQc & ca-CondMat & US powergrid \\
		\hline
		RW-HDP & 62.26 & 1124.51 & 504.16 & 1262.18 & 235.46 \\
		SIP2-LDA & 279.95 & 1664.80 & 2902.81 & 41920.72 & 7197.49 \\
		\hline
	\end{tabular}
	\caption{perplexity}
	\label{table:perplexity}
\end{table}
                           
\begin{figure}[H]
	\centering
	\includegraphics[scale=0.6]{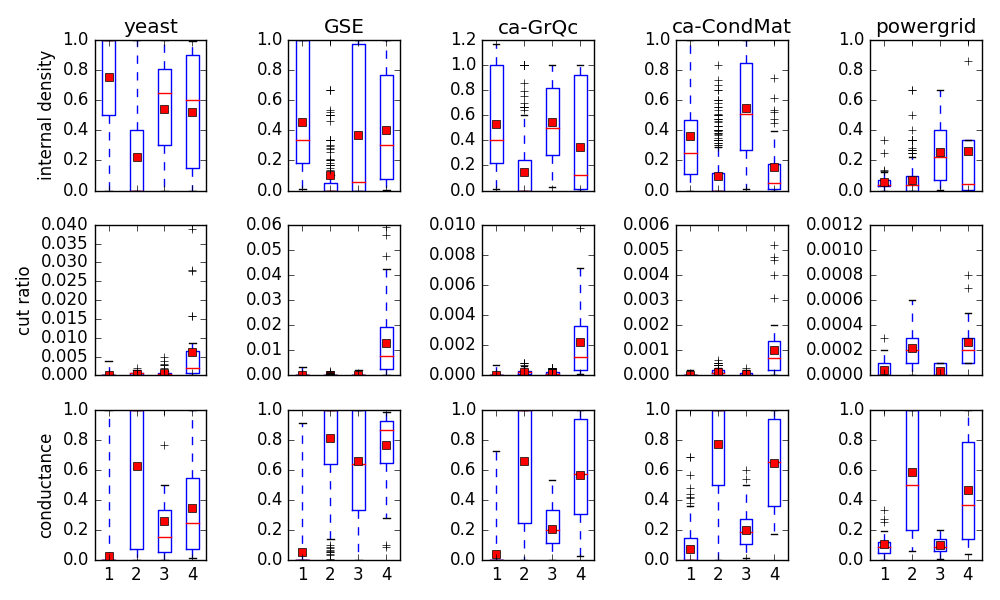}
	\caption{Boxplots of internal density, cut ratio and conductance. 1: RW-HDP, 2: SIP2-LDA, 3: Walktrap, 4: BCD.}
\end{figure}

Our model outperforms other models almost in every aspect and on every dataset, except that the modularity of our model on the GSE network is slightly smaller than that of Walktrap and the densities of our model on the US powergrid and the ca-CondMat network are not the biggest among all models. \\

RW-HDP outperforms SIP2-LDA because random walk is a much better way of collecting information compared to the social interaction profile. Also, HDP topic model is more powerful than LDA, as the former is a nonparametric Bayesian model. Both RW-HDP and Walktrap utilize random walks but perhaps it is because the transition probability between nodes within $t$ steps is not the best way to capture nodes similarities that makes Walktrap less successful in terms of model performances. Also, Walktrap views the community detection as an optimization problem. It merges two communities according to Ward's method which minimizes the mean of squared distances between each node and its community. To get the best partition, the algorithm cuts the hierarchical tree at where it has the largest modularity, which could result in less satisfying results on other metrics. RW-HDP, on the contrary, is a probabilistic model, which may be the reason why it is more flexible. As for BCD, though it is also a nonparametric Bayesian model, hierarchically, it has a relatively shallow depth, which could be the reason why it lacks the power to reveal the network structure.

\section{Conclusion and Future Works}
Network community detection has been studied intensively by researchers from various disciplines. In this paper, we present the RW-HDP model for this task. As the name suggested, the two pillars for RW-HDP are random walk and Hierarchical Dirichlet Process topic model. We first conduct random walks on the network and treat them as documents. Later, we fit the Hierarchical Dirichlet Process topic model to reveal community structure. As HDP is a nonparametric model, our method enables us to find the number of communities automatically. Our work is a new endeavor in nonparametric Bayesian modeling in networks. It borrows ideas from some novel previous works and outperforms them. \\

The choices of hyperparameters are based on intuitive rules at the current stage. A more sophisticated mathematical formulation leaves as one of our future works. Also, our model can be improved in many ways. One possible future work is to allow overlapping community detection. Currently, we assume each node only belongs to a unique community, the one gives the largest conditional probability. We may relax this assumption and assign a community to a node if its corresponding conditional probability $p(z=k|w)$ exceeds some threshold, which would allow overlapping communities to exist. Finally, we believe by changing HDP topic model to other topic models, we are able to find various types of community structures in networks, such as hierarchical community structure. Such interesting applications of topic models on social networks are yet to be explored in the future.

\end{document}